\documentclass[12pt,aps,eqsecnum,amsfonts,amsmath]{article}
\usepackage{amssymb}
\begin{document}
\newcommand{\ZZ}{{\mathbb Z}} \newcommand{\RR}{{\mathbb R}}
\newcommand{\NN}{{\mathbb N}} \newcommand{\comp}{{\scriptstyle \circ}} 
\newcommand{\loc}{_{\rm local}} \newcommand{\glo}{_{\rm global}}
\newcommand{\ee}{\hbox{e}} \newcommand{\Tr}{\hbox{Tr}}
\newcommand{\Ad}{\hbox{Ad}} 
\newcommand{\eins}{\mathbf 1} \newcommand{\QED}{\hspace*{\fill}Q.E.D.} 
\renewcommand{\today}{}

\title{\vskip-15mm \bf Fusion rules for the continuum sectors \\ 
of the Virasoro algebra with $c=1$}
\author{Karl-Henning Rehren\thanks{Electronic address: 
{\tt rehren@theorie.physik.uni-goettingen.de}}  \\ and \\
Hilmar R. Tuneke\thanks{Electronic address: 
{\tt tuneke@theorie.physik.uni-goettingen.de}} \\[3mm]
Institut f\"ur  Theoretische Physik, Universit\"at  G\"ottingen,\\
37073 G\"ottingen, Germany} 

\maketitle

\begin{abstract}
The Virasoro algebra with $c=1$ has a continuum of superselection
sectors characterized by the ground state energy $h\geq 0$. Only the
discrete subset of sectors with $h=s^2$, $s\in\frac12\NN_0$, arises by
restriction of representations of the $SU(2)$ current algebra at level
$k=1$. The remaining continuum of sectors is obtained with the help of
(localized) homomorphisms into the current algebra. The fusion product
of continuum sectors with discrete sectors is computed. A new method
of determining the sector of a state is used.
\\[2mm] PACS 11.10.Cd, 11.25.Hf 
\end{abstract}

\section{Introduction}
``Fusion rules'' describe the product of two superselection charges
and the decomposition of the product into irreducible charges. They
thus constitute an important characteristics for the charge 
structure of a quantum field theory. 

The general definition of the composition of charges (``DHR product'')
was first given in \cite{DHRprod}. In two-dimensional conformal
quantum field theory, other notions of fusion \cite{BPZ,N} became more
popular, but every evidence shows \cite{FRS,W} that these describe
the same abstract charge structure.  

The actual computation of the fusion rules in concrete models is
in general a difficult task, and almost always relies on some specific
apriori knowledge. If the QFT at hand is the fixpoint subalgebra of
another QFT with respect to a compact gauge group, then harmonic
analysis determines the composition law for those sectors which appear
in the decomposition of the vacuum sector of the larger algebra
\cite{DHRfix}. The fusion rules then follow the composition of the
representations of the gauge group. In low-dimensional theories, a
gauge group is in general not present, but in favorable cases, 
modular transformation properties \cite{V} or ``null vectors''
\cite{BPZ,N} can be exploited.

In the present letter we treat a model where the standard strategies are not
applicable: the chiral stress-energy tensor of a 1+1-dimensional
conformal quantum field theory with $c=1$. (A chiral field can be
treated like a ``one-dimensional QFT''.) Its algebra $A$ is the fixpoint 
algebra of the chiral $SU(2)$, level $k=1$, current algebra $B$ with
respect to its global $SU(2)$ symmetry \cite{Fk,KHR}, and the
positive-energy representations of the current algebra contain a
discrete series of superselection sectors of $A$. But besides the
discrete series there is a continuum of further sectors which do not
arise by restriction from $B$. These sectors have no ``null vectors''
and hence infinite asymptotic dimension \cite{KHR}, so that the
Verlinde formula or Nahm's prescription are not applicable.      

We adopt a method due to Fredenhagen \cite{F} for the computation of
the fusion rules: A charged state $\omega$ is described by a positive
map $\chi$ of the algebra into itself such that 
$$\omega=\omega_0\comp\chi$$
where $\omega_0$ is the vacuum state. The correspondence between states 
and positive maps is 1:1 provided the charge is strictly localized.
This yields a product of states defined by
$$\omega_1 \times \omega_2 := \omega_0\comp\chi_1\comp\chi_2. $$
The GNS representation $\pi_{\omega_1 \times \omega_2}$ is always a 
subrepresentation of the DHR product of GNS representations
$\pi_{\omega_1} \times \pi_{\omega_2}$ \cite{F}, and is expected to
exhaust it as the positive maps vary within their equivalence class. 

For two states $\omega_1$ and $\omega_2$ belonging to the discrete and
continuous sectors, respectively, we shall determine (by a new
method) the sectors to which the product states belong. 

\section{Fusion rules for \hbox{$c=1$}}
The superselection sectors of the stress-energy tensor with $c=1$ are 
uniquely determined by their ground state energy $h\geq 0$ for the
conformal Hamiltonian $L_0$. The sectors $[h=s^2]$ with $s\in\NN_0$,   
arise as subrepresentations of the vacuum representation of the
$SU(2)$ current algebra $B$, and those with $s\in\NN_0+\frac12$ arise
in the spin-$\frac12$ representation of $B$. Those with
$h\notin(\frac12\ZZ)^2$ constitute the continuum. For each of these
representations, the partition function is well known \cite{part}: 
$$\Tr \exp(-\beta \pi_h(L_0))=\left\{\begin{array}{lcl} t^h p(t) & \hbox{if}&
    h\notin(\frac12\ZZ)^2, \\ (t^{s^2}-t^{(s+1)^2}) p(t) &
\hbox{if}& h=s^2, \quad s \in\frac12\NN_0,\end{array}\right.$$
where $t=\ee^{-\beta}$ and $p(t)=\prod_n(1-t^n)^{-1}$. 

The positive maps describing the charged states are of the form
(cf.\ Lemma 2.1)
$$\chi = \mu\comp\alpha_g\vert_A.$$
Here $g$ is a smooth $SU(2)$ valued function, and $\alpha_g$ the
automorphism of the current algebra $B$ induced from the local gauge
transformation (Bogolyubov automorphism) of the underlying chiral
fermion doublet, 
$$\psi_i(x) \mapsto \sum_j\psi_j(x)g_{ji}(x).$$
$\mu=\int d\mu(k)\,\gamma_k$ is the average over the global gauge group 
$SU(2)$ acting by automorphisms $\gamma_k$. Since $\mu$ is a positive
map of $B$ onto $A$, $\chi_g$ is a positive map of $A$ onto $A$. 

The induced action of $\alpha_g$ on the currents 
$j(f)\equiv\sum j^a(f_a)=\int:\psi(x) f(x)\psi(x)^*:dx$ (with an $su(2)$
valued test function $f(x)=\sum f_a(x) T^a$) is explicitly computed as 
$$\alpha_g(j(f))=j(gfg^{-1})-\frac i{2\pi}\int\Tr(fg^{-1}\partial g)\eins,$$
and its restriction to the Sugawara stress-energy tensor 
$T=\frac\pi 3 \sum g_{ab}:j^aj^b:$ is 
$$\alpha_g(T(f)) = T(f) -ij(f\partial gg^{-1}) -\frac 1{4\pi}\int
f\,\Tr(\partial gg^{-1}\partial gg^{-1})\eins.$$
The central terms arise, of course, from normal ordering.
To be specific, we choose the functions
$$g_q(x)=\pmatrix{\exp(iq\lambda(x)) & 0\cr 0 & \exp(-iq\lambda(x)) }$$
where $\lambda(x)=-i \log\frac{1+ix}{1-ix}$ interpolates between
$\lambda(-\infty)=-\pi$ and $\lambda(+\infty)=+\pi$, and $q$ is a real
parameter whose role as a charge will be exhibited in Lemma 2.1. \footnote{It
  appears that one could also use the embedding $T=\pi :jj:$ of $A$
  into a $U(1)$ current algebra $C$. The problem would be that the  
  conditional expectation $\mu$ which takes the homomorphisms
  $\alpha_g: A\to C$ back onto $A$ in order to obtain 
  $\chi = \mu\comp\alpha_q\vert_A$ is not explicitly known in that case.}  

At this point, we have to distinguish the quasilocal algebras $A\loc$ and 
$B\loc$ generated by field operators smeared with test functions, and the
global algebras $A\glo$ and $B\glo$ generated by field operators
smeared with ``admissible'' functions which are test functions up to
polynomials of order $2(d-1)$ where $d$ is the scaling dimension. It is
well known \cite{LM} that the fields as distributions extend to these
enlarged test function spaces, so that 
$$L_n=\frac12\int(1-ix)^{1-n}(1+ix)^{1+n}T(x)dx 
\quad\hbox{and}\quad
Q^a_n=\int(1-ix)^{-n}(1+ix)^{n} j^a(x)dx$$
are defined as closed unbounded operators. The specific automorphisms
$\alpha_q\equiv\alpha_{g_q}$ extend to the operators
$Q^3_n\in B\glo$ and $L_n\in A\glo$:
$$\alpha_q(Q^3_n)=Q^3_n + q\delta_{n,0}, \qquad\alpha_q(L_n)=L_n + 
2q Q^3_n + q^2 \delta_{n,0}\qquad(q\in\RR),$$ 
but they extend to $Q^\pm_n\in B\glo$ only if $q\in\frac12\ZZ$: 
$$\alpha_q(Q^\pm_n)=Q^\pm_{n\pm2q}\qquad(q\in\frac12\ZZ).$$ 
(Our basis of $SU(2)$ and hence of the fields $j^a$ is such that
$[Q^+_n,Q^-_m]=2Q^3_{n+m}+n\delta_{n+m,0}$,
$[Q^3_n,Q^\pm_m]=Q^\pm_{n+m}$, $[Q^3_n,Q^3_m]=\frac 12 n\delta_{n+m,0}$.) 

Our first Lemma establishes the relation between the parameters $q$ and $h$:

\vskip2mm{\bf 2.1.~Lemma:} The state
$\omega_q\equiv\omega_0\comp\chi_q \equiv
\omega_0\comp\mu\comp\alpha_q\vert_A = \omega_0\comp\alpha_q\vert_A$
is a ground state in the irreducible sector $[h=q^2]$.\vskip2mm

{\it Proof:} Since the operators $L_n$ and $Q^3_n$ ($n\geq 0$) annihilate 
the vacuum, $\alpha_q(L_n)$ annihilate the vacuum for $n>0$ and 
$\alpha_q(L_0)$ has eigenvalue $q^2$. It follows that %the GNS vector for 
$\omega_q$ is a ground state for $L_0$ with ground state energy $q^2$. \QED

Thus, in order to compute the fusion rules $[h_1]\times[h_2]$ (where
$h_i=q_i^2$) one has to determine the GNS representation for the
product state
$$\omega_{q_1}\times\omega_{q_2} 
= \omega_0\comp\alpha_{q_1}\comp\mu\comp\alpha_{q_2}\vert_A =
\int_{SU(2)}d\mu(k)\; \omega_0\comp\alpha_{q_1}\comp\gamma_k\comp
\alpha_{q_2}\vert_A .$$ 
This state is a continuous mixture of states 
$\omega_k \equiv\omega_0\comp\alpha_k$ induced by the homomorphisms
$$\alpha_k\equiv\alpha_{q_1}\comp\gamma_k\comp\alpha_{q_2}\vert_A$$ 
of $A\loc$ into $B\loc$. (We suppress the explicit reference to the
involved charges $q_1$ and $q_2$.) These homomorphisms extend to $A\glo$ for
generic $k\in SU(2)$ only if $q_1\in \frac12\ZZ$, as can be seen from
the above transformation formulae. The following argument is more
physical: If one evaluates $\omega_k(T(f)^2)$ for test functions $f$,
then one finds that the contributions from the current two-point
functions diverge for generic $q$ as $f$ is replaced by the function
$\frac12(1+x^2)$. Hence the operator $L_0=\frac12\int (1+x^2)T(x)dx$
has a finite expectation value but infinite variance in these states.  

This is why we shall restrict ourselves to the case $q_1\in\frac12\ZZ$.
Since $q$ and $-q$ give rise to the same sector $[h=q^2]$, we shall even
assume $q_1\in\frac12\NN_0$.

Now we exploit the fact that $\gamma_k$ is implemented by a unitary
operator in $B\glo$ of the form $U(k)=\exp(i\sum \kappa_a Q^a_0)$ on
which $\alpha_{q_1}$ is well defined. Hence
$$\alpha_k=\Ad(V(k))\comp \alpha_{q_1}\comp\alpha_{q_2}\vert_A =
\Ad(V(k))\comp \alpha_{q_1+q_2}\vert_A$$
with $V(k) = \alpha_{q_1}(U(k))= \exp(i\sum\kappa_a\alpha_{q_1}(Q^a_0))$.
It is more convenient to express $U(k)$ in the form
$$U(k)=\exp(i\frac{k_2^*}{k_1}Q^-_0)k_1^{2Q^3_0}\exp(i\frac{k_2}{k_1}Q^+_0)
\qquad \hbox{for} \qquad k=\pmatrix{ k_1 & ik_2 \cr ik_2^* & k_1^* }.$$ 
($k_1^{2Q^3_0}$ is well defined since $2Q^3_0$ has integer spectrum.)
Application of $\alpha_{q_1}$ yields 
$$V(k)= k_1^{2q_1}\exp(i\frac{k_2^*}{k_1}Q^-_{-2q_1})k_1^{2Q^3_0}
\exp(i\frac{k_2}{k_1}Q^+_{+2q_1}).$$

{\bf 2.2.~Lemma:} The product state $\omega_{q_1}\times\omega_{q_2}$ 
is a convex integral over states $\omega_0\comp\alpha_k$, $k\in SU(2)$. 
Each state $\omega_0\comp\alpha_k$ on $A$ is a finite convex sum  
$$\omega_0\comp\alpha_k = \sum_{\nu=0}^{2q_1}\pmatrix{2q_1 \cr \nu}
\vert k_1\vert^{2(2q_1-\nu)}\vert k_2\vert^{2\nu} \;
\omega_{q_1,q_2}^{(\nu)}$$ 
of states 
$$\omega_{q_1,q_2}^{(\nu)}(\,\cdot\,) = \frac{(2q_1-\nu)!}{(2q_1)!\nu!}\;
((Q^-_{-2q_1})^\nu\Omega,
\alpha_{q_1+q_2}(\,\cdot\,)(Q^-_{-2q_1})^\nu\Omega).$$
Since only the weights depend on the group element $k\in SU(2)$, the
product state $\omega_{q_1}\times\omega_{q_2}$ is a finite convex sum
of the same states $\omega_{q_1,q_2}^{(\nu)}$. \vskip2mm

{\it Proof:} The first statement just summarizes the precedent discussion. 
We have $\omega_0\comp\alpha_k = 
(V(k)^*\Omega,\alpha_{q_1+q_2}(\,\cdot\,)V(k)^*\Omega)$, and 
$V(k)^*\Omega =(k_1^*)^{2q_1}\exp(-i\frac{k_2^*}{k_1^*}Q^-_{-2q_1})\Omega$ 
because $Q^a_n$ annihilate the vacuum for $n\geq 0$ (remember our choice
$q_1\in\frac12\NN_0$). The power series expansion of the exponential
yields vectors $(Q^-_{-2q_1})^\nu\Omega$ with energy $2q_1\nu$ and
Cartan charge (the eigenvalue of $Q^3_0$) $C=-\nu$. These vectors
vanish for $\nu>2q_1$ because the vacuum Hilbert space $H$ of $B$ does not
contain vectors with energy less than $C^2$. This fact is read off the
following expression \cite{part} for the partition function for the
vacuum representation: 
$$\Tr \exp(-\beta L_0-\eta Q^3_0)=\sum_{j\in\NN_0}\sum_{m=-j}^j
z^m (t^{j^2}-t^{(j+1)^2})p(t)\qquad(z=\ee^{-\eta},t=\ee^{-\beta})$$
in which the power of $t$ is always at least the square of the power
of $z$. Since $\alpha_{q_1+q_2}(L_n)$ does not change the Cartan
charge $C$, the vectors $(Q^-_{-2q_1})^\nu\Omega$ have only diagonal
matrix elements for $\alpha_{q_1+q_2}(A)$, showing the convex
decomposition. The proper normalization of the states
$\omega_{q_1,q_2}^{(\nu)}(1)=1$ can be checked recursively in $\nu$. 
\QED  

The problem has thus been reduced to the determination of the GNS 
representations $\pi_{q_1,q_2}^{(\nu)}$ for the states
$\omega_{q_1,q_2}^{(\nu)}$. One can easily compute that these states are
eigenstates of $L_0$ with energy $(q_1+q_2)^2-2\nu q_2$, but they are not 
ground states in general. It is therefore not possible to determine the
sectors directly via their ground state energies. Instead, it turns out to
be possible to compute the partition function for the representations
induced by these states. This is our main result. 

\vskip2mm{\bf 2.3.~Proposition:} Let $q_1\in\frac12\NN_0$. If
$q_2\notin\frac12\ZZ$, then $\pi_{q_1,q_2}^{(\nu)}$ is irreducible and
belongs to the sector $[h=(q_1+q_2-\nu)^2]$. If $q_2\in\frac12\ZZ$,
then $\pi_{q_1,q_2}^{(\nu)}$ is a direct sum of sectors from the set
$\{[h=s^2] : s\in\vert q_1+q_2-\nu\vert+\NN_0\}$. \vskip2mm

{\it Proof:} The vector $(Q^-_{-2q_1})^\nu\Omega$ has Cartan charge
$C=-\nu$. This value is not changed by application of
$\alpha_{q_1+q_2}(L_n)$, hence $\pi_{q_1,q_2}^{(\nu)}$ is a
subrepresentation of the representation $\alpha_{q_1+q_2}$ on the
subspace $H_{C=-\nu}=P_{-\nu}H$ of Cartan charge $-\nu$ in the vacuum
representation of $B$. The partition function for the latter
representation is   
$$\Tr\, P_{{-\nu}}\exp(-\beta\alpha_{q_1+q_2}(L_0))=
\ee^{-(q_1+q_2)^2\beta}\cdot 
\Tr\, P_{{-\nu}}\exp(-\beta L_0 - 2(q_1+q_2)\beta Q^3_0).$$
From the previous expression for the vacuum partition function, we
obtain 
$$\Tr\, P_{{-\nu}}\exp(-\beta L_0 - \eta Q^3_0)=z^{-\nu}t^{\nu^2} p(t)
\qquad(z=\ee^{-\eta},t=\ee^{-\beta})$$
by collecting the terms $z^{-\nu}$, and hence
$$\Tr\, P_{{-\nu}}\exp(-\beta\alpha_{q_1+q_2}(L_0))=
t^{(q_1+q_2-\nu)^2}p(t).$$
If $q_1+q_2-\nu\notin\frac12\ZZ$, then this is the partition function
of the irreducible sector $[h=(q_1+q_2-\nu)^2]$. Hence
$\alpha_{q_1+q_2}(A)$ acts irreducibly on $H_{C=-\nu}$, and must
coincide with its subrepresentation $\pi_{q_1,q_2}^{(\nu)}$. If on the
other hand $q_1+q_2-\nu\in\frac12\ZZ$, then the above equals the sum
of the partition functions $(t^{s^2}-t^{(s+1)^2})p(t)$ of the sectors
$[h=s^2]$ with $s\in\vert q_1+q_2-\nu\vert+\NN_0$. Thus
$\pi_{q_1,q_2}^{(\nu)}$ is the direct sum of a subset of these sectors. 
\QED 

As mentioned in the introduction, the product of states, computed
here, might accidentally not exhaust the DHR product. But this
degeneracy disappears if the positive map $\chi_2$ is perturbed by the
adjoint action of some isometry $a\in A$. We note that the argument
leading to Prop.\ 2.3 is in fact stable if $\chi_{q_2}$ is replaced by 
$\Ad(a^*)\comp\chi_{q_2}$. Namely, because $a$ is $SU(2)$-invariant, one has 
$\Ad(a^*)\comp\gamma_k\comp\alpha_{q_2} = \Ad(U(k)a^*)\comp\alpha_{q_2}$, 
so it is sufficient to replace in the above argument the vectors
$(Q^-_{-2q_1})^\nu\Omega$ by the perturbed vectors
$\alpha_{q_1}(a)(Q^-_{-2q_1})^\nu\Omega$ which still belong to 
$H_{C=-\nu}$. In the case $q_2\notin\frac12\ZZ$, the perturbed GNS
representation $\pi_{q_1,q_2}^{(\nu)}$ will still belong to the
irreducible sector $[h=(q_1+q_2-\nu)^2]$. 

Thus, combining Lemma 2.2 with the Proposition, we obtain

\vskip2mm{\bf 2.4.~Corollary:} Let $q_1\in\frac12\NN_0$ and 
$q_2\in\RR\setminus\frac12\ZZ$. The fusion rules for the sectors
$[h_i=q_i^2]$ are 
$$[h_1]\times[h_2]=\bigoplus_{\nu=0}^{2q_1}\;[h^{(\nu)}] \qquad\hbox{with}
\qquad h^{(\nu)}=(q_1+q_2-\nu)^2.$$

\section{Comments}
We have studied the decomposition into irreducibles of the product
of sectors (``fusion rules'') for the chiral stress-energy tensor with
$c=1$. We succeeded to compute the fusion rules for two sectors with ground
state energies $h_i$ where $[h_1]$ is a special sector,
$h_1\in(\frac12\NN_0)^2$, and $[h_2]$ belongs to the continuum of
sectors, $h_2 \in\RR_+\setminus(\frac12\NN_0)^2$, Cor.\ 2.4. This
result was not accessible by the prevailing methods for the
computation of fusion rules. The case where both sectors belong to the
continuum should in principle also be studied with the present method, 
but becomes technically very intricate. 

When both sectors $[h_i]$ are special, we would have 
expected $SU(2)$-like fusion rules \cite{DHRfix} since the special
sectors $[h=s^2]$, $s\in\frac12\NN_0$, arise by restriction of the
vacuum and spin-$\frac12$ representations of $B$ to the fixpoint
algebra $A$ on the subspaces of $SU(2)$ charge $s$. This is, however,
not reproduced by Prop.\ 2.3 and Lemma 2.2: Although the unperturbed 
states $\omega_{q_1,q_2}^{(\nu)}$ have finite energy and hence only 
finitely many of the possible sectors according to Prop.\ 2.3 really 
contribute to them, this limitation will disappear if $\chi_{q_2}$ is 
perturbed as described above. Moreover, if $h_i=s_i^2$ with $0<s_2<s_1$, 
the sectors $[h=s^2]$ with $0\leq s<\vert s_1-s_2\vert$ should not occur 
according to $SU(2)$, while they are not excluded by Prop.\ 2.3, and
are really found to be present by more explicit computations.  

This state of affairs has a simple explanation: For $q\in\frac12\ZZ$, 
the positive maps $\chi_q$ transfer not only the $SU(2)$ charge
$s=\vert q\vert$ but in fact, as explained below, a mixture of all charges 
$s\in\vert q\vert+\NN_0$. These admixtures are not seen if evaluated in
the vacuum state (Lemma 2.1), but become visible if evaluated in a
generic state of $A$, e.g., upon perturbation of $\chi_q$. The 
product states $\omega_0\comp\chi_{q_1}\comp\chi_{q_2}$, too, are
sensitive to admixtures to $\chi_{q_2}$, which accounts for the
presence of ``too many'' sectors contributing to the fusion rules as
inferred from Lemma 2.2 and Prop.\ 2.3.   

Let us explain why $\chi_{q}$ is capable of transferring the ``wrong''
charges if $q\in\frac12\ZZ$, but not if $q\notin\frac12\ZZ$, and why 
this is not in conflict with the statement in \cite{F} that the 
correspondence between states and positive maps is 1:1. The argument is 
very similar to the one in the proof of Prop.\ 2.3. If $\chi_q$ is 
evaluated in some perturbed state $\omega=(a\Omega,\,\cdot\,a\Omega)$
with $a\in A$, we have $\omega\comp\chi_q=\omega\comp\alpha_q$ since $a$
and $\omega_0$ are $SU(2)$ invariant. Thus the GNS representation
$\pi_\omega$ for $\omega$ is a subrepresentation of the representation
$\alpha_q$ on the subspace $H_{C=0}=P_0H$ of Cartan charge $C=0$ in the
vacuum representation of $B$ (to which $a\Omega$ belongs). The
partition function for this representation has been computed above
(putting $q_1=0,\nu=0,q_2=q$):
$$\Tr\, P_{0}\exp(-\beta\alpha_{q}(L_0))= t^{q^2}p(t).$$
This is the character of the irreducible representation $[h=q^2]$ if
$q\notin\frac12\ZZ$, but is the sum of infinitely many irreducible
characters for $[h=s^2]$, $s\in\vert q\vert+\NN_0$, if $q\in\frac12\ZZ$.

By testing with suitable operators $a\in A\glo$, one finds that the
``wrong'' sectors are indeed present. Remember that the 1:1
correspondence between states and positive maps requires that the 
charge is strictly localized, while the automorphisms $\alpha_q$ in
our analysis are only asymptotically localized (the derivative
$\partial g_q(x)$ vanishes  asymptotically). Of course our choice for
$\alpha_q$ was dictated by the simplicity of the transformation
formulae for $L_n$ and $Q^a_n$. The unpleasant feature of the wrong
sectors is the price for that simplification. 

The fusion rules in Cor.\ 2.4 are not affected by this complication.

This work is based on the Diploma Thesis of the second author \cite{T}.

\end{document}